\documentclass[sigconf,10pt]{acmart}
\usepackage{physics}
\usepackage{amsmath}
\usepackage[utf8]{inputenc}
\usepackage{natbib}
\usepackage{graphicx}
\usepackage{subcaption}
\usepackage{pbox}
\usepackage{amsmath}
\usepackage{amsfonts}
\usepackage{flushend}
\usepackage{graphicx}
\usepackage{tabularx}
\usepackage{tabularx}
\usepackage{adjustbox}
\usepackage{algorithm}
\usepackage{algpseudocode}
\def \sys {\textit{DeepCell}}

\usepackage{booktabs} 
\usepackage{comment}
\usepackage{xcolor}
\usepackage{multirow}

\setcopyright{acmcopyright}
\copyrightyear{2021}
\acmYear{2021}
\acmDOI{10.1145/1122445.1122456}

\acmConference[SIGSPATIAL'21]{SIGSPATIAL '21: ACM 27th Annual International Conference On Mobile Computing And Networking}{October 25-29, 2021}{New Orleans, LA}
\acmBooktitle{New Orleans '21: ACM 27th Annual International Conference On Mobile Computing And Networking,
  October 25-29, 2021, New Orleans, LA}
\acmPrice{15.00}
\acmISBN{978-1-4503-XXXX-X/18/06}

\begin{document}

\title{DeepCell: A Ubiquitous Accurate Provider-side Cellular-based Localization}


\author{Ahmed Shokry}
\affiliation{%
  \institution{Pennsylvania State University}
  \city{PA}
  \country{USA}}
\email{ahmed.shokry@psu.edu}

\author{Moustafa Youssef}
\affiliation{%
  \institution{American University in Cairo}
  \city{Cairo}
  \country{Egypt}}
\email{moustafa-youssef@aucegypt.edu}

\renewcommand{\shortauthors}{Shokry and Youssef}

\begin{abstract}
	Although outdoor localization is already available to the general public and businesses through the wide spread use of the GPS, it is not supported by low-end phones, requires a direct line of sight to satellites and can drain phone battery quickly. The current fingerprinting solutions can provide high-accuracy localization but are based on the client side. This limits their ubiquitous deployment and accuracy.

	In this paper, we introduce \sys{}: a provider-side fingerprinting localization system that can provide high accuracy localization for any cell phone. To build its fingerprint,
	\sys{} leverages the unlabeled cellular measurements recorded by the cellular provider while opportunistically synchronizing with selected client devices to get location labels. The fingerprint is then used to train a deep neural network model that is harnessed for localization. To achieve this goal, \sys{} need to address a number of challenges including using unlabeled data from the provider side, handling noise and sparsity, scaling the data to large areas, and finally providing enough data that is required for training deep models without overhead.

	Evaluation of \sys{} in a typical realistic environment shows that it can achieve a consistent median accuracy of 29m. This accuracy outperforms the state-of-the-art client-based cellular-based systems by more than 75.4\%. In addition, the same accuracy is extended to low-end phones.
\end{abstract}



\maketitle
\section{Introduction}
\label{introduction}

In recent years, much attention is being paid to developing mobile applications that require outdoor position information. These applications include navigation systems, location-based advertisements, location-based social networks, and location-based emergency services.
\cite{elhamshary2014checkinside,shankar2012crowds,zheng2011location}.
The Global Positioning System (GPS) \cite{hofmann2012global} is considered the most commonly used localization system due to its accuracy. However, GPS requires a direct line-of-sight to satellites, which limits its accuracy and availability in urban areas and bad weather conditions outdoors 
\cite{cui2003autonomous}. Moreover, GPS incurs an unacceptable energy consumption, which severely limits the use of smartphones and reduces the battery lifetime \cite{gaonkar2008micro}. To alleviate the high-energy requirement of the GPS, several techniques are proposed \cite{youssef2010gac,constandache2010compacc,jurdak2010adaptive,paek2010energy}.
These techniques focus on how to improve GPS by combining a duty-cycling of the GPS with mobile sensors data to conserve energy and prolong battery life.

Alternatively, numerous solutions have been proposed to address the problem of high energy consumption of GPS. WiFi-based outdoor localization systems have been proposed, e.g. 
\cite{cheng2005accuracy,lamarca2005place,Skyhook,thiagarajan2009vtrack}. They leverage the Received Signal Strength (RSS) overheard from WiFi access points deployed in buildings along the roads as a metric for localization. However, WiFi signals are not available in many road areas, e.g. on highways and rural areas. Moreover, WiFi signals outside buildings are usually weak which limits their ability to differentiate distinct locations. To further enhance the localization accuracy and reduce the energy consumption, new systems leverage smartphones low-energy augmented sensors, such as the compass, gyroscope, and accelerometer, for localization e.g. \cite{aly2013dejavu,constandache2010towards,ofstad2008aampl,azizyan2009surroundsense,arthi2010localization,light}. These systems can provide high localization accuracy with low-energy consumption. However, similar to GPS and WiFi-based systems, these systems are not available in low-end phones that are still used throughout the world, limiting their ubiquitousness.

To enable ubiquitous localization, a number of localization systems have been proposed based on the cellular signals that can be scanned with virtually zero extra power besides the standard phone operations\cite{ibrahim2012cellsense,elnahrawy2007adding,li2008angle,varshavsky2005gsm}. These techniques use propagation models or fingerprinting for estimating user location \cite{vo2016survey}.
The propagation models techniques, e.g. \cite{elnahrawy2007adding,li2008angle,elbakly2016robust,elbakly2015calibration}, try to obtain a relation between signal strength and distance. Due to signal reflection, interference, and attenuation, propagation models techniques achieve limited localization accuracy outdoors\cite{weiss2003accuracy,ghaboosi2011geometry}. On the other hand, fingerprinting techniques, e.g. \cite{ibrahim2012cellsense,ibrahim2010cellsense,ergen2014rssi}, usually work in two phases. Offline phase is an initial training phase (i.e., calibration) during which the received signal strength (RSS) measurements from the different cell-towers in the area of interest are recorded from the users side (i.e. client side) at known locations. Online phase where RSS measurements from the overheard cell-towers at an unknown location are matched against the fingerprint database to determine the best location match.
To build the fingerprint from client side,
current techniques use a scanning program that records the GPS location label and RSS information for the cell-tower the mobile is connected
to as well as other neighboring cell-towers information in the 2G GSM networks. According to the GSM standards \cite{ibrahim2010cellsense}, each cell phone can receive signals from at most seven cell-towers, one of them is the current cell-tower the phone is connected to  (serving cell-tower) and the others are the neighboring cell-towers. However, many cell phones, even advanced ones, do not provide API to get access to the neighboring cell-towers’ information. This problem appears also for other cellular networks e.g. 3G UMTS where neighboring cell-towers’ information is also not available. Moreover, the old generations of cell phones do not have a scanning program to measure the RSS which is used for localization. This brings up a challenge of providing accurate localization for different network types over all kinds of phones. 

In this paper, we introduce \sys{}: a provider-side fin-
gerprinting localization system that can provide high accuracy
localization for any cell phone. Specifically, \sys{} leverages the unlabeled cellular data that is recorded by the cellular provider based on users' phones activities to build the fingerprint. \sys{} then trains a deep model to automatically capture the unique signatures of the different cell-towers at different fingerprint locations.  
To achieve \sys{} goals, new challenges still need to be addressed including how to build the fingerprint from the provider unlabeled data without overhead, handling noise and sparsity in provider data, scaling the data to large areas, and finally providing enough data that is required for deep learning models without overhead.

Evaluation of \sys{} on different phones in typical area show that \sys{} can achieve a consistent median localization error of 29m which is better than the state-of-the-art cellular localization techniques by more than 75.4\%. This comes with zero extra energy consumption compared to the normal phone operation. This highlights the use of our system for ubiquitous accurate cellular localization. 

The rest of the paper is organized as follows:
Section~\ref{provider}
describes the recorded provider data. Section~\ref{overview}
presents our system architecture. Section~\ref{details} gives the details
of the \sys{} system followed by its
evaluation in Section~\ref{evaluation}. Section~\ref{history} discusses related
work. Finally, we conclude the paper in Section~\ref{conclusion}.

\section{Provider celluar data}
\label{provider}

The provider cellular data contains records for the different phone numbers. It is event-based data. The provider collects data only for a specific phone events such as when the phone changes the cell-tower it is connected to. Each cellular measurements record is a tuple contains the following items:

\begin{itemize}
	\item Event type.
	
	\item Timestamp in Greenwich Mean Time (GMT) during which the event happens.
	
	\item Phone ID which is the phone number. 
	
	\item Set of active cell-towers (active cells) which is the set of most probable cell-towers the phone connected or may connect to. For each cell-tower, the cell-tower ID, the Radio Network Controller (RNC), and the received signal strength (RSS) information is provided.
	
	\item Neighboring cell-towers information. For each cell-tower, the cell-tower ID, RNC, and the received signal strength (RSS) information are recorded.
	
\end{itemize}

\noindent The provider RSS information is measured from the control channel which can provide a robust RSS information as the transmit power is constant. 
Also, the provider records the active cells which makes its fingerprint more informative than building the fingerprint from users side. 

On the other hand, the provider celular data is unlabeled which means that it needs to be geo-tagged with ground truth locations in order to be used to build the fingerprint. Finally, as the provider data is event-based, the resulting fingerprint is noisy and sparse in both time and space. 

\section{System Architecture}
\label{overview}

\begin{figure}[t!]
	\centering
	\includegraphics[width=0.99\columnwidth]{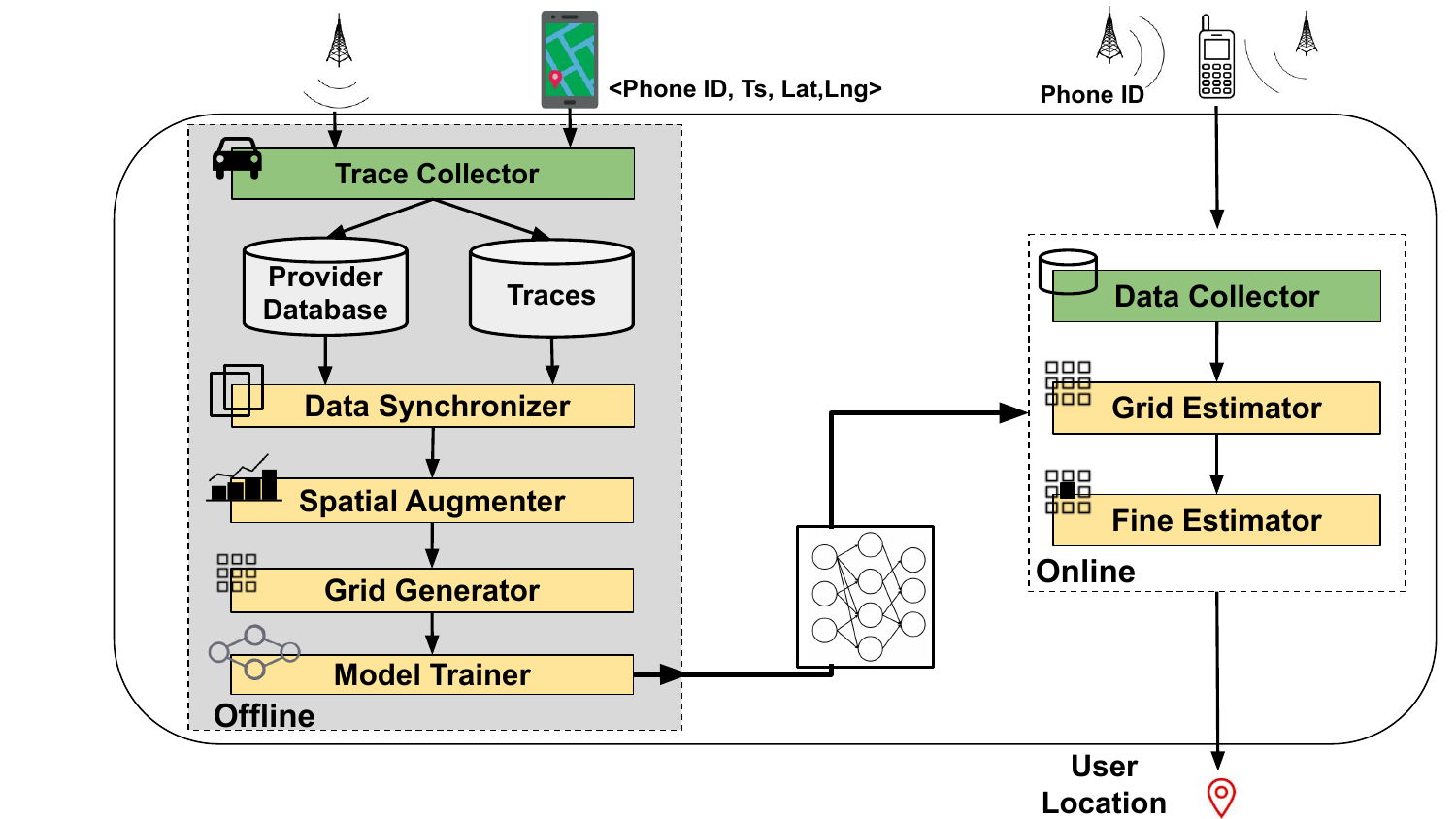}
	\caption{\sys{} system architecture.}
	\label{arcitecture}
\end{figure}

In this section, we describe \sys{} main components and give an overview about its operation. Figure~\ref{arcitecture} shows the system architecture. \sys{} works in two phases: offline training phase and online tracking phase. During the offline phase, GPS traces are collected by war-drivers in a crowd-sourcing mannar using the Trace Collector module. The module is deployed on smart-phones to collect phone ID (SIM card number), phone date and time, and GPS location information. Meanwhile, the module enables the provider to build the cellular database for the different phones' events that are used while collecting the GPS traces. The GPS traces database and the provider database are then forwarded to the Data Synchronizer module.
\par
The Data Synchronizer module is used to synchronize the GPS traces from clients with the provider cellular data. Specifically, it assigns GPS location labels from the GPS traces database to the unlabeled cellular measurements records from the provider database. The labeled cellular data are then fed to the Spatial Augmenter module.   
\par
The Spatial Augmenter module is used to handle the noise and sparsity in the training data (i.e. labeled cellular data). It increases the number of training samples by predicting the unknown cellular measurements for a new set of locations. The training samples after the spatial augmention are then passed to the Grid Generator module.
\par
Traditionally \cite{youssef2005horus,bahl2000radar}, to build the fingerprint, the data collectors have to stop and stay at the different fingerprint locations for enough time, typically a few minutes, to gather the RSS samples. This increases the fingerprint construction overhead \cite{shokry2017tale}, especially in large areas, which is the norm outdoors. To reduce this overhead, the Grid Generator module super-imposes a virtual grid over the area of interest and the geo-tagged cellular measurements collected inside each cell are used as training samples to construct the deep model for this cell. 
\par
The Model Trainer module trains a deep neural network model using the labeled cellular measurements records that are associated with the grid information. The input to the model is the cellular measurements from the different cell towers heard in the area of interest. The output is a probability distribution over the different grid cells in the area of interest. Special precaution are used to avoid over-training as well as to increase the model robustness. Note that the model is incrementally and dynamically updated as more training samples become available to the provider database, allowing \sys{} to maintain a fresh fingerprint all the time.
\par
During the online tracking phase, a user standing at an unknown location. Given the phone ID, the provider can fetch the cellular measurements of the user's phone using the Data Collector module and forward them to the Grid Estimator module. The Grid Estimator module uses the trained deep network to calculate the probability distribution of each grid cell and fuses this information using the Fine Estimator module to estimate the final user location.
%

\begin{table}[!t]
	\centering
	\caption{ Table of notations}
	\label{tab:notations}
	\scalebox{0.8}{
		
		\begin{tabular}{|c|l|}\hline
			Symbol & Description \\ \hline \hline
			$D_u$ & Provider unlabeled cellular data. \\ \hline
			$D_s$ & Provider labeled cellular data (sparse). \\ \hline
			$D_d$ & Provider labeled cellular data after spatial augmentation (dense). \\ \hline
			$\mathbb{G}$ & Universe of grid cells in the area of interest (i.e virtual grid).\\ \hline
			$K$ & Number of grid cells. \\ \hline
			$M$ & Total number of cell towers in the environment. \\ \hline
			$N_s$ & Total number of training samples. \\ \hline
			$N_a$ & Total number of synthesised training samples by spatial augmenter. \\ \hline
			$l_i$ & Location label (latitude and longitude) where $1 \leq i \leq N_a$.\\ \hline
			$r_i$ & Synthesised training sample $|r_i|$ = 2$M$ where $1 \leq i \leq N_a$.\\ \hline			
			
			$x_i$ & Model input (i.e training sample $|x_i|$ = 2$M$) where $1 \leq i \leq N_s$. \\ \hline
			$y_i$ & Model output (i.e logits $|y_i|$ = $K$) for corresponding input $x_i$.  \\ \hline
			$y_{ij}$ & Output score that scan $x_i$ is at grid cell $j$. \\ \hline
			$p(y_{ij})$ & Probability that scan $x_i$ is at grid cell $j$.\\ \hline
			$P(y_i)$ & Probability distribution for different grid cells ($|P|$ = $K$)).\\ \hline
			$L_i$ & \pbox{9cm}{One-hot encoded vector present the probability distribution for different grid cells in the training phase ($|L_i|$ = $K$)).} \\ \hline
			$x$ & Online RSS sample/scan vector $|x|=2M$.\\ \hline
			$P(g|x)$ & Probability of receiving a signal vector $x$ at grid cell $g \in \mathbb{G}$.\\ \hline
			$g^*$ & Estimated grid cell.\\ \hline
			$l^*$ & Center of mass of $g^*$.\\ \hline
			\hline
			
		\end{tabular}
	}
\end{table}

\section{System Details}
\label{details}

In this section, we present the \sys{} system details including the offline model training phase and the online user tracking phase. We also describe how the system handles different practical challenges. Table~\ref{tab:notations} summarizes the notations used in this section.
\subsection{The Offline Model Training Phase}
During this phase, the deep network model is trained taking into account a number of challenges including: 
building the fingerprint from the provider unlabeled cellular data without overhead, handling noise and sparsity in provider data, scaling the data to large areas, providing enough data that is required for training deep models without overhead, training the deep model while avoiding over-training and finally increasing model robustness. 

\subsubsection{The data synchronizer}
This module is responsible for synchronizing (i.e. matching) the GPS location labels from the traces database with the cellular measurements records from the provider unlabeled database. The provider cellular data $D_u$ contains a timestamp in Greenwich Mean Time (GMT). On the other hand, the GPS scans have the phone time and date. This motivates us to time-synchronize the cellular measurements with the GPS location labels. To this end, we convert the phone timestamp to be in GMT. We consider a GPS label matches a cellular measurements record if the time difference between them is less than a certain thereshold. 
\begin{equation}
\vert T_{measurements} - T_{label} \vert \leq \Psi
\end{equation} 
We choose the matching tolerance threshold $\Psi$ to be 10ms.

\subsubsection{The spatial augmenter}
\label{sec:data_aug}

In practice, the quality and the size of the training data influence the deep learning classification accuracy. As the provider data is an event-based data, the resulting cellular data is sparse. Hence, the goal of this module is to handle the noise and sparsity in the provider labeled cellular data $D_s$. It increases the number of training samples by predicting the unknown fingerprint values (i.e. cellular measurements) for the different locations that are not in the provider cellular samples ($D_s$). The new constructed fingerprint values contains not only the RSS from cell-towers in the environment, but also a bit associated with each cell-tower in the environment to indicate wheather or not the cell-tower belongs to the active cells. 

We leverage a Gaussian Process (GP) to capture the relation between the provider cellular measurements records and the location labels assuming that the relation between a cellular measurements record $r_i$ at location $l_i$ is as follows: 
\begin{equation}
r_i = f(l_i) + \epsilon_i
\end{equation}
Where $\epsilon_i$ is an irreducible error. 

To approximate the mapping function $f(.)$, the Gaussian Process
starts with a prior distribution and updates this distribution as data samples are observed from the provider data $D_s$, producing the posterior distribution over functions.
These distributions are represented non-parametrically, in terms of the training samples.
Specifically, the function $f(.)$ is distributed as a GP with mean function $m(.)$ and a positive definite covariance function $k(.)$,

\begin{equation}
f \sim \mathcal{GP}(m,k)
\end{equation} 

The covariance function (i.e. kernel) between two locations $l_p$ and $l_q$ is defined as a squared exponential or Gaussian Kernel:
\begin{equation}
k(l_p,l_q) = \exp(-\frac{1}{2l^2}\vert l_p - l_q \vert^2)
\end{equation}

where $l$ is is the length scale that determines how strongly the correlation between points drops off. The key idea is that if  $l_p$ and $l_q$ are deemed by the kernel to be similar, then we expect the output of the function at those points to be similar, too.

The GP provides a way for extending the provider sparse fingerprint $D_s$ to a more dense one $D_d$ as it gives the fingerprint values (i.e. cellular measurements) at any arbitrary location based on the assumed model. To ensure an accurate approximation for the fingerprint values, we obtain the fingerprint values only for locations on the paths of the GPS traces.

\subsubsection{The grid generator} 
\begin{figure}[t!]
	\centering
	\includegraphics[width=0.8\columnwidth]{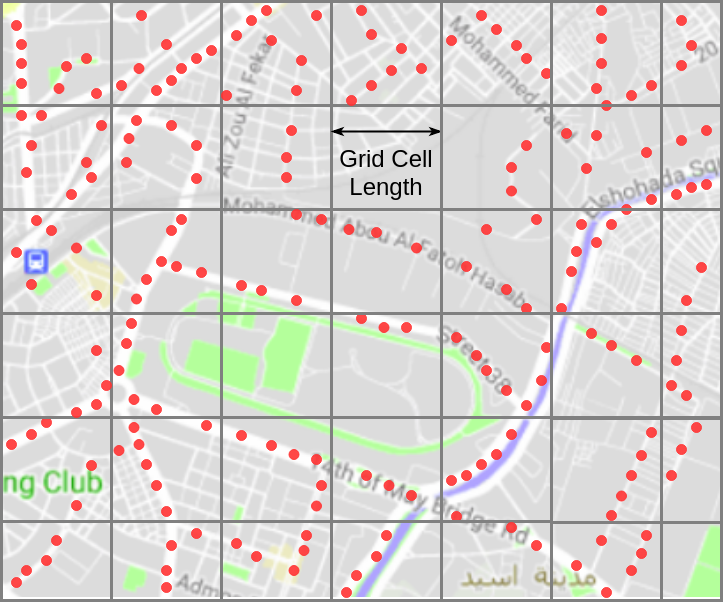}
	\caption{The gridding approach. The area of interest is superimposed by equally-sized square cells. Red points present the training samples.}
	\label{fig:grid} 
\end{figure}

To remove the requirement of the data collectors standing still at different fingerprint locations for a certain time, the Grid Generator module divides the area of interest into a virtual grid (Figure~\ref{fig:grid}). Each training sample in the form of \textit{(location label, cellular measurements)} is mapped to a specific cell based on its coordinates. All the samples within a virtual grid cell are used as training samples for this cell signature. The cell length is a parameter that presents a trade-off between the localization accuracy and the computational-complexity as we quantify in Section~\ref{evaluation}.
\par
The gridding approach can handle multiple scalability challenges. It allows the data to be collected while the war-drivers/users are moving naturally in their life, without requiring them to stand still at the different fingerprint locations. Moreover, the gridding approach can control the model complexity/accuracy through controlling the number of grid cells in the area of interest.

\subsubsection{The Model trainer}
\begin{figure}[!t]\center
	\includegraphics[width=0.9\linewidth]{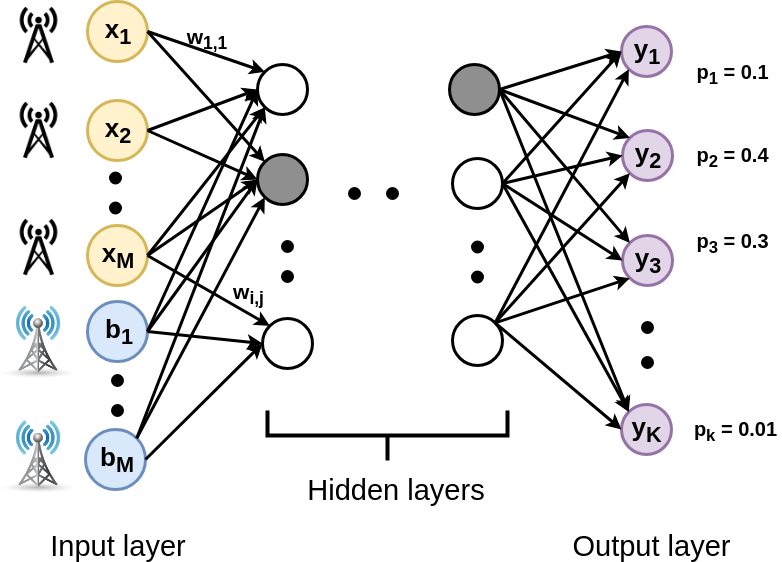}
	\caption{Network structure. The input is the RSS coming from different cell towers in the environment with a bit corresponds to each cell tower indicates wheather or not the cell-tower is in the active cells (features). The output is the probability distribution for different grid cells in the area of interest. Grey-shaded neurons represent examples of nodes that have been temporary dropped-off to increase the model robustness and avoid over-training.}
	\label{fig:structure}
\end{figure}

The target of this module is to train the deep neural network model using the \textit{spatially-augmented} fingerprint samples dataset constructed by the \textit{Spatial Augmenter} module. Figure~\ref{fig:structure} presents our deep network model structure. The input to the classifier is the cellular measurements coming from $M$ cell-towers that cover the \textit{entire} area of interest. Specifically, each fingerprint sample input consists of the RSS coming from the $M$ cell towers in the area of interest and a bit corresponds to each cell-tower tells us wheather or not the cell-tower is in the active cells. If a cell-tower is not heard in a scan, \sys{} sets its corresponding RSS measurement/active cell bit to zero. The output is the grid cell probability distribution, i.e. the probability that the input sample/scan belongs to each cell in the area of interest.

\sys{} employs a multinomial logistic model~\cite{hosmer2013applied} which acts as a non-linear function whose independent variables (i.e network input) is a set of real-valued cellular measurements coming from different cell towers deployed in the area of interest. The dependent variables (i.e network output) is a probability distribution over the virtual grid cells.
Since each input scan in the augmented data belongs to only one grid cell, the output probability distribution is presented as one-hot encoded vector with one at the grid cell the input training sample is in and zero otherwise.
The model can infer the probability distribution of different grid cells through learning the feature vector of the different training examples. 
\par
More formally, assume that we have $N_s$ training samples present our spatially-augmented fingerprint training set collected through corwd-sensing. 
Each record $x_i$ where $ 1\leq i \leq N_s$ contains $M$ real value features $(x_{i1},x_{i2},..x_{iM})$ that present the RSS information coming from $M$ cell towers covering the entire area of interest and $M$ bit features $(b_{i1},b_{i2},..b_{iM})$ that present wheather or not the cell-tower is in the active cells. The corresponding discrete outcomes (i.e logits) for the input $x_i$ is $y_i = (y_{i1},y_{i2},..,y_{iK})$ present the score for each grid cell of the possible $K$ grid cells to be the estimated cell. \sys{} employs the Softmax function to convert the logit score $y_{ij}$ (for sample $i$ to be at grid cell $j$) to a probability. The score probability is calculated by the Softmax function as follows
\begin{equation}
p(y_{ij}) = \frac{e^{y_{ij}}}{\sum_{j=1}^{K}e^{y_{ij}}}
\end{equation}
where 
\begin{equation}
\sum_{j=1}^{K} p(y_{ij}) = 1
\end{equation}
During the training phase, the expected (i.e. ground-truth probability) vector $P(y_i) = [ p(y_{i1}) , p(y_{i2}) ...   p(y_{iK})]$ is presented as one-hot vector ($L_i$) which has one for the entry corresponding to the correct grid cell and zeros for other grid cells. 
To train the model we employ a stochastic gradient descent optimizer that attempts to minimize the average cross-entropy distance between the estimated output vector probability $P(y_i)$ and the one-hot vector $L_i$ over the entire training dataset. 
The loss function is defined as:
\begin{equation}
\mathrm{loss} = \frac{1}{N_s} \sum_{i=1}^{N_s} D(P(y_i), L_i)
\end{equation}
where $P(y_i)$ is the network output for input $x_i$, computed using the Softmax function, $L_i$ is the one-hot encoded vector for sample $i$, and $D(P(y_i), L_i)$ is the cross-entropy distance function defined as:
\begin{equation}
D(P(y_i), L_i) = - \sum_{j=1}^{K} l_{ij} \log(p_(y_{ij}))
\end{equation}
\par
The multinomial logistic classifier learns the relationship, captured by the model weights, between the input features (i.e input cellular measurements variables $x_i$) and the grid cells probability distribution $P(y_i)$, so that the outcome of a new cellular measurements vector can be correctly predicted for a new data point with unknown outcome.
\par
To mitigate the effect of the inherent noise in the input training samples and further increase the model robustness, \sys{} employs a drop-out regularization technique during training \cite{srivastava2014dropout}. The basic idea is to randomly drop neurons from the network during training (shaded nodes in Figure~\ref{fig:structure}). The temporary removed neurons no longer contribute to the activation of downstream neurons in the forward pass. Similarly,  the weight update process is not applied to them in the backward pass. This prevents the network from over-fitting the training data. 

\subsection{The Online User Tracking Phase}
During the online user tracking phase, \sys{} \textit{depends only on the phone ID} which is the SIM card number as an input. The Data Collector module fetches the cellular measurements that corresponds to the phone ID of the user standing at unknown location.

To estimate the user location, the Grid Estimator module first estimates the discrete grid cell $g^* \in \mathbb{G}$ in which the user exists given the cellular measurements vector $x=(x_1, x_2,...x_M,b_1, b_2,...b_M)$. 
To this end, \sys{} leverages the trained deep model. Specifically, the input vector is then fed to the deep network model and 
the grid cell with the maximum probability is selected as the the estimated grid cell. More formally, 

\begin{equation}
g^* = \underset{g \in \mathbb{G}}{\operatorname{argmax}}[P(g|x)]
\end{equation}

The center of mass of $g^*$ represents a coarse-grained location estimate of the user location. To further refine the estimated position in the continuous space, the Fine Estimator module estimates user location $l^*$ as the center of mass of all grid-cells, weighted by their respective probability:
\begin{equation}
l^* =\sum_{g \in \mathbb{G}} g \times p_g
\end{equation}
where $\mathbb{G}$ is the universe of grid cells in the area of interest and $p_g$ is the corresponding estimated probability at the output layer of the model.


\section{Performance evaluation} \label{evaluation}
In this section, we evaluate \sys{} performance. We start describing our testbed. Then, we study the effect of different parameters on \sys{} performance. Finally, we quantify the overall system performance and compare it to the state-of-the-art techniques. 

\subsection{Data collection}
\begin{figure}[t!]
	\centering
	\includegraphics[width=0.8\columnwidth]{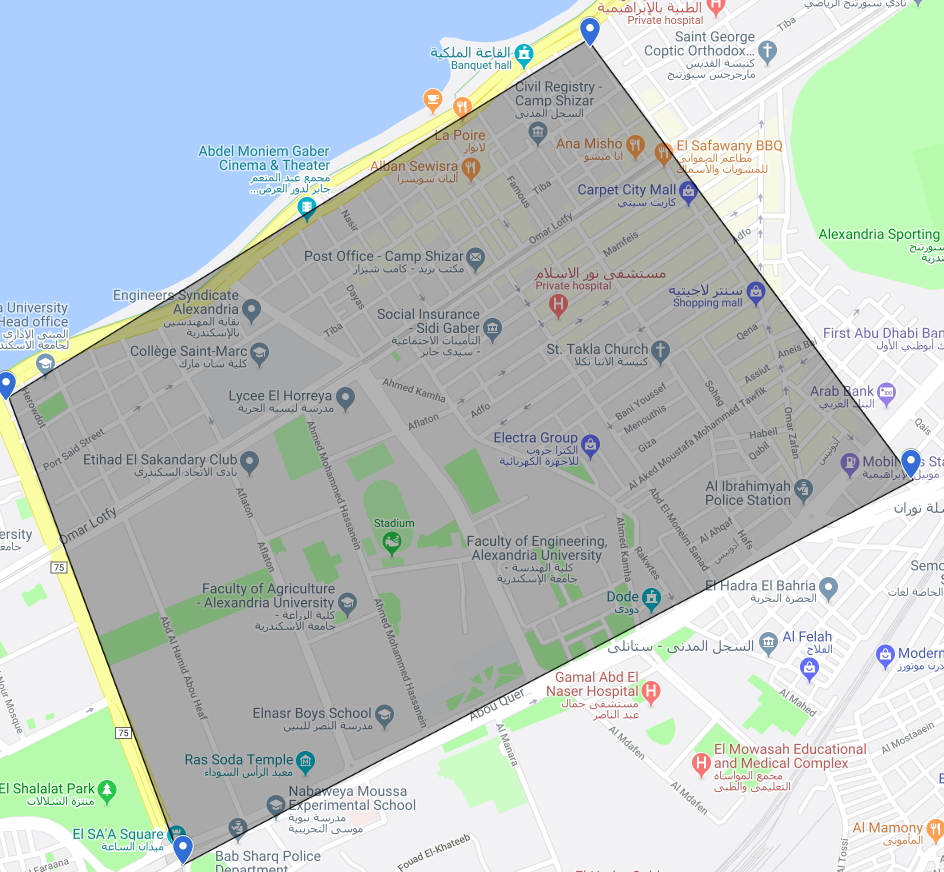}
	\caption{Our testbed.}
	\label{fig:testbed} 
\end{figure}

Our testbed covers a 2.04$\text{Km}^2$ in a typycal urban area in Alexandria, Egypt. Number of samples collected by the cellular provider is 2013 sample. Number of cell-towers in the environment is 42 cell-tower. Figure \ref{fig:testbed} shows our testbed.

GPS traces are collected by war-drivers using different devices including HTC One X9 and Motorola Moto G5 plus phones among others.
Drivers' devices have GPS receivers, which we used for collecting the ground-truth location labels. The deployed collector software collects GPS ground-truth locations and timestamps. The collectors also are connected to the internet using mobile data which enables the cellular provider to collect the unlabeled cellular data.

For training the deep network model, We tried different network structures. The best structure contains three hidden layers with 256, 128, and 64 neurons. This structure produces a consistent accuracy. We study the effect of the training epochs and learning rate on the system performance later in this section.

\subsection{Effect of \sys{} Parameters}
In this section, we study the effect of different parameters on \sys{} 
performance including the effect of the different fingerprint models, changing grid spacing, changing the number of augmented samples used in training, cell towers density, number of training epochs, and learning rate. Table \ref{default} summarizes the system parameters used throughout the experiments and their default values.

\begin{table}[!t]
	\centering
	\caption{Parameters and their default values}
	\label{default}
	\scalebox{0.8}{
		\begin{tabular}{|l||l|p{2cm}|}\hline
			\textbf{Parameter} & \textbf{Range} & \textbf{Default value}\\ \hline \hline
			Fingerprint model & \pbox{10cm}{RSS only,\\RSS with active cells,\\ Spatial augmentation,\\ All}  & All\\ \hline
			Grid cell length ($G_s$) & 50-1500m  & 100m\\ \hline
			Number of augmented samples ($N_s$) & 0-1000 & 1000 \\ \hline
			Number of training epochs ($N_e$) & 50-500 & 500 \\ \hline
			Learning rate ($\alpha$) & 0.01-0.0001  & 0.0001\\ \hline
			Cell towers density ($D_s$) &  25\%-100\% & 100\% \\ \hline
		\end{tabular}
	}
\end{table}

\begin{figure*}[!t]
	\minipage{0.32\textwidth}
	\includegraphics[width=\linewidth]{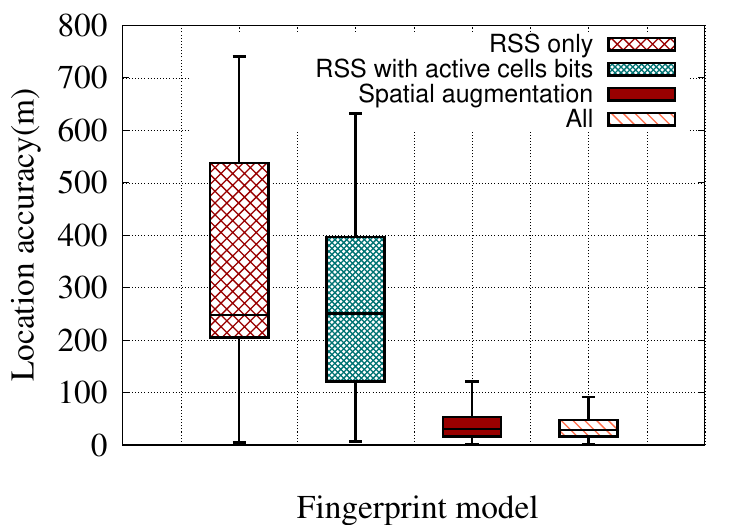}
	\caption{Effect of different fingerprint models on the localization accuracy.}
	\label{fig:fmodels}
	\endminipage\hfill
	\minipage{0.32\textwidth}
	\includegraphics[width=\linewidth]{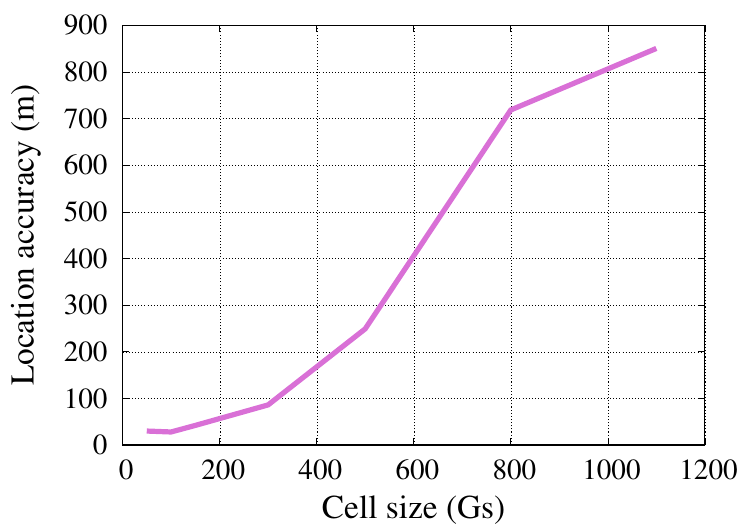}
	\caption{Effect of changing grid cell length on the localization accuracy.}
	\label{fig:cellsize}
	\endminipage\hfill
	\minipage{0.32\textwidth}
	\includegraphics[width=\linewidth]{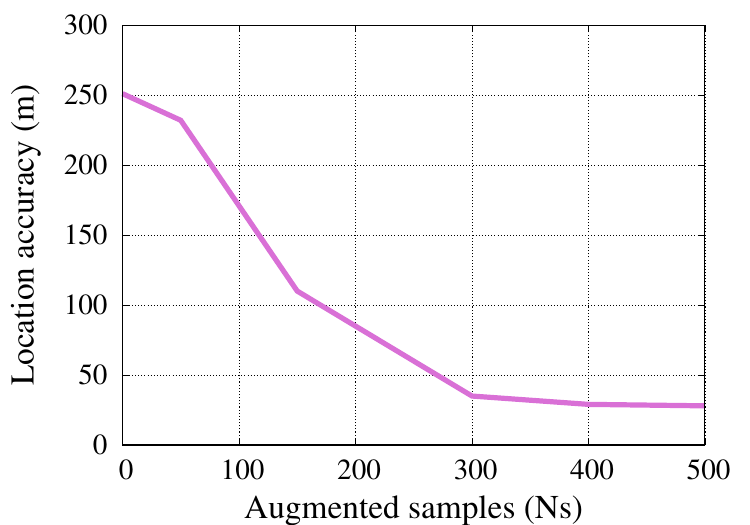}
	\caption{Effect of changing number of augmented training samples on the localization accuracy.}
	\label{fig:trainingsamples}
	\endminipage\hfill
\end{figure*}
\begin{figure*}[!t]
	\minipage{0.32\textwidth}
	\includegraphics[width=\linewidth]{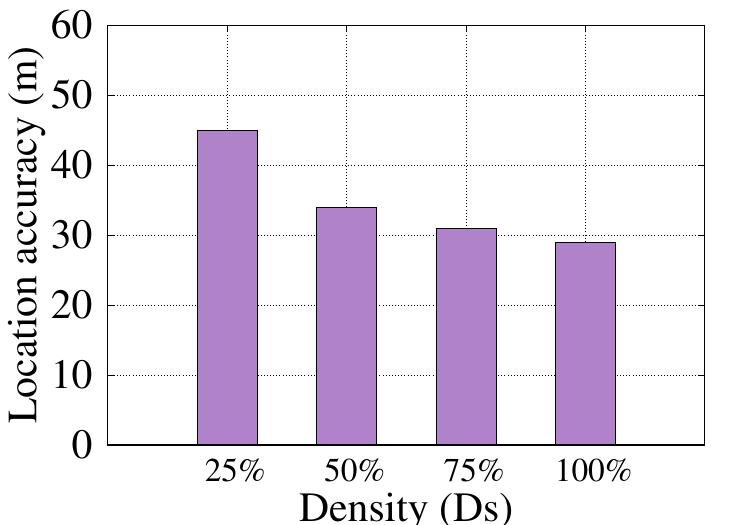}
	\caption{Effect of changing cell towers density on the localization accuracy.}
	\label{fig:density}
	\endminipage\hfill
	\minipage{0.32\textwidth}
	\includegraphics[width=\linewidth]{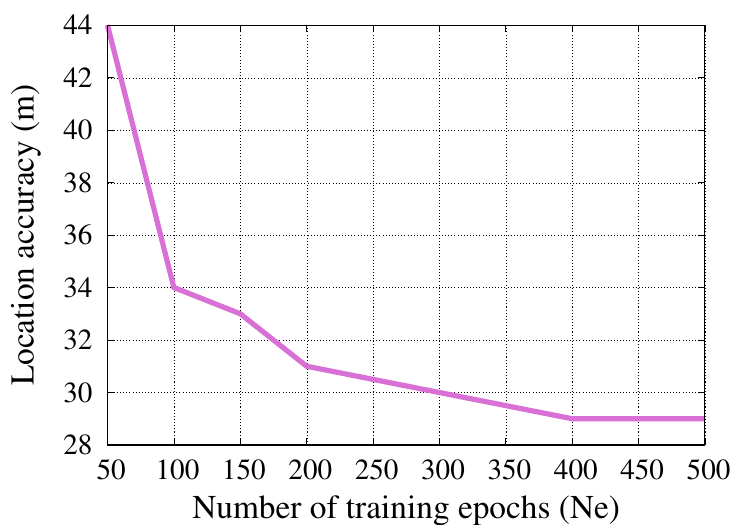}
	\caption{Effect of changing number of training epochs on the localization accuracy.}
	\label{fig:epochs}
	\endminipage\hfill
	\minipage{0.32\textwidth}
	\includegraphics[width=\linewidth]{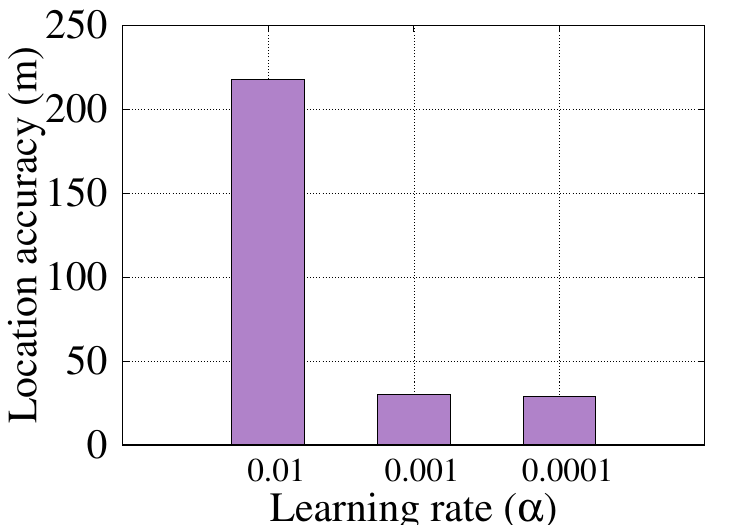}
	\caption{Effect of changing learning rate on the localization accuracy.}
	\label{fig:learningrate}
	\endminipage\hfill
\end{figure*}

\subsubsection{Effect of fingerprint models}
Figure~\ref{fig:fmodels} shows the effect of different fingerprint models on the system accuracy. The first model uses the received signal strength coming from the different cell-towers only as an input (i.e.features) for training the neural network. The second model uses the RSS from the different cell-towers with a bit associated with each cell-tower indicates wheather or not the cell-tower is in the active cells. The third model applies the spatial augmentation technique before training the neural network. Finally, the fourth model uses the RSS with active cells bits and then augments the model with new samples generated using the spatial augmentation technique.  

The figure shows that both the active cells information and the spatial augmentation can enhance system accuracy compared to training without them. The figure confirms that the spatial augmentation can significantly improve system accuracy, not only by increasing the number of training samples, but also by implicitly handling the inherent noise in the input GPS location labels and cellular data values. Using both active cells information with spatial augmentation leads to the best accuracy with higher accuracy. 

\subsubsection{Effect of virtual grid spacing ($G_s$)}
Figure \ref{fig:cellsize} shows the effect of changing the grid cell length on the system median localization accuracy.
The figure shows that as the grid cell length increases, the system accuracy degrades. This can be explained by observing that as the grid cell length increases the grid cells cover larger areas, leading to coarser-grained accuracy. The optimal value for the grid spacing is at $G_s=100m$.

\subsubsection{Effect of number of augmented samples used in training ($N_s$)}
Figure~\ref{fig:trainingsamples} shows the effect of increasing the number of augmented samples on the system accuracy. The figure shows that increasing number of augmented training samples increases the system accuracy due to the increase of the quality of the learned model. \sys{} performance saturates after about 500 samples. This can be further reduced using the data augmentation module as quantified before.  

\subsubsection{Effect of cell towers density ($D_s$)}
Figure~\ref{fig:density} shows the effect of cell towers density percentage on the system accuracy. To reduce the cell tower density, we remove a percentage of the cell-towers from the input feature vector with their active cells bits as indicated in the figure. The figure shows that increasing the cell towers density leads to better accuracy due to the addition of new features (the RSS from the new towers with their active cells bits). These features lead to better  discrimination between the different grid cells, which in turn results in a better accuracy. \sys{} maintains its accuracy even with 75\% of the typical density.

\subsubsection{Effect of number of training epochs ($N_e$)}
Figure~\ref{fig:epochs} shows the effect of changing the number of training epochs on the system accuracy. This parameter controls the number of times the system iterates over the entire training dataset to update the weights.
The figure shows that increasing number of training epochs leads to a better localization accuracy. However, increasing number of training epochs over 400 has no effect on the system accuracy.

\subsubsection{Effect of changing learning rate ($\alpha$)}
Figure~\ref{fig:learningrate} shows the effect of changing the learning rate on the localization accuracy. This parameter enables the learning optimizer to control the weights update in the direction of the gradient during the optimization progress.
The figure shows that lowering the learning rate to ($\alpha = 0.0001$), while performing a fair number of training epochs, makes the training more reliable. This is because lowering the learning rate makes the optimizer step towards the minimum slowly which prevents overshooting it. Note that the network training is performed during the offline phase, which allows us to increase the number of epochs for better training without overhead. 

\subsection{Comparison with other systems}
In this section, we compare \sys{} performance to the state-of-the-art deep learning-based technique (i.e the \textit{DeepLoc} system \cite{shokry2018deeploc}) which provides the best performance among the current outdoor cellular localization systems. \textit{DeepLoc} builds the fingerprint from the user side then it uses it for localization. In particular, it stores the geo-tagged RSS information coming from each cell-tower in the area of interest at each grid cell. This fingerprint is then used to train a deep neural network model for localization. During the online tracking phase, DeepLoc uses the RSS measurements that are recorded by the user phone to localize it.

\begin{figure}[!t]
	\centering
	\includegraphics[width=0.95\columnwidth]{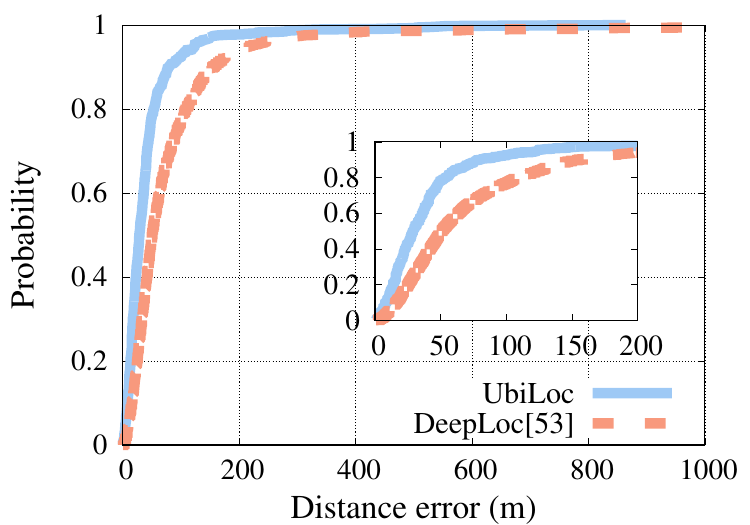}
	\caption{CDF of localization error.}
	\label{fig:cdfs}
\end{figure}

Figure~\ref{fig:cdfs} shows the CDF of distance error for both systems. Table~\ref{comparison} further summarizes the results. The figure shows that our system, \sys{}, can achieve a better localization accuracy than DeepLoc system by more than 75.4\%. Moreover, \sys{} enhances all the other quantiles. This highlights the use of \sys{} system as ubiquitous accurate localization system. 

\vspace{0.8cm}

\section{Related work}
\label{history}

\sys{} leverages the provider unlabeled cellular measurements from the different cell towers to train a deep neural network model to achieve ubiquitous accurate outdoor localization. In this section, we discuss the localization techniques literature and how they differ from \sys{}. 

\subsection{GPS-based Techniques}
The Global Positioning System (GPS) \cite{hofmann2012global} is considered as the most commonly used outdoor localization technique. However, it suffers from a number of drawbacks. It incurs an unacceptable energy consumption. Also, it requires a direct line-of-sight with the satellite, limiting its coverage in areas; e.g. with tall buildings and inside tunnels; and even in bad weather conditions \cite{gu2021effect,cui2003autonomous}. To reduce the high energy consumption of GPS, a number of techniques have been proposed that combine other techniques, e.g. map matching or inertial sensors \cite{aly2015semmatch}, with a duty cycled GPS \cite{youssef2010gac,constandache2009enloc}. Those techniques trade the accuracy with low power consumption. In addition, GPS and inertial sensors are still not available in a variety of low-end phones existing throughout the world, limiting their ubiquitousness.

\sys{}, on the other hand, is an energy-efficient technique. It can provide a highly accurate localization in different environments. In addition, the cellular RSS is available in all cell phones, providing ubiquitous coverage.

\begin{table*}[!t]
	\centering
	\caption{Comparison between DeepLoc and \sys{}}
	\label{comparison}
	\begin{tabular}{|l|l|l|l|l|l|}\hline
		Accuracy & Minimum &\pbox{10cm}{$25^{th}$\\percentile }& \pbox{10cm}{$50^{th}$\\percentile}&\pbox{10cm}{$75^{th}$\\percentile}& Maximum\\ \hline \hline		
		\textbf{\sys{}} & \textbf{0.3} & \textbf{15.8}  & \textbf{29} & \textbf{47} & \textbf{863}\\ \hline
		DeepLoc\cite{shokry2018deeploc} & 0.7 (-133\%)  & 29 (-83.5\%) &  50.89 (-75.4\%) & 94 (-100\%) & 4101 (-375\%)\\ \hline \hline
	\end{tabular}
\end{table*}

\subsection{Sensors-based Techniques}
The techniques in this category leverage WiFi APs \cite{cheng2005accuracy,lamarca2005place,Skyhook,rekimoto2006placeengine,yoshida2005locky} or smart-phones augmented sensors \cite{shokry2020dynamicslam,aly2013dejavu,azizyan2009surroundsense,constandache2010towards,ofstad2008aampl,abdelnasser2016semanticslam,wang2012no} to localize users. The basic idea behind these techniques is to use a WiFi fingerprint or landmarks detected by the different phone sensors to localize the phone. These systems can provide an accurate location with a low-energy consumption. However, WiFi-based techniques suffer from coverage problems as WiFi signals are weak outdoors which limits their ability to differentiate between distinct locations. Furthermore, these systems can not provide the estimated location in areas with no deployed APs, such as on highways and rural areas. In addition, both WiFi and augmented sensors-based techniques are not ubiquitously available on all cell phones. 

\sys{}, on the other hand, takes the advantages of ubiquitous coverage of cellular networks and is available on all cell phones.

\subsection{Cellular-based Techniques}
These techniques leverage either the Cell-ID \cite{paek2011energy} or the RSS from cell-towers to provide ubiquitous localization \cite{ibrahim2012cellsense,constandache2009enloc,ibrahim2010cellsense,laoudias2018survey}. The Cell-ID based techniques eatimate user's position from the strongest cell-towers heard by the user. Hence, they provide a coarse-grained accuracy. 
The RSS-based techniques leverage the cellular RSS to localize the user. These techniques employ probabilistic models, such as Baysian-based models as in \cite{ibrahim2012cellsense,ibrahim2011hidden,ibrahim2010cellsense}, to localize users. They usually assume that the RSS coming from the different cell-towers are \textit{independent} to to make the problem mathematically tractable and avoid the curse of dimensionality \cite{bishop2006pattern} which limits their ability to capture the mutual relation between the RSS from the different cell towers, which in turn lowers their accuracy.

\sys{}, on the other hand, employs a deep-model to capture the joint distribution of input features, which leads to significantly better performance than the traditional probabilistic models.

\subsection{Deep Models-based Techniques}
Recently, a number of deep learning based fingerprinting techniques have been proposed indoors \cite{rizk2018cellindeep,Ibrahim2018} and outdoors \cite{shokry2018deeploc,rizk2019effectiveness}. The indoor localization systems collect the fingerprint training samples manually, using stationary receivers which need to stand for a few minutes at each discrete fingerprint locations. This high overhead process does not scale to the large areas which is the norm outdoor.

The current outdoor localization systems collect the fingerprint samples from the users side in 2G GSM network. However, many cell phones, even advanced ones, do not provide API to get access to the neighboring cell-towers’ information. In addition, many cell phones do not have API to get neighboring cell-towers' information in 3G UMTS networks. Furthermore, the old generations of cell phones can not provide the RSS information even for the cell-tower the phone is connected to. This brings up a new challenge of providing accurate localization for different network types over all kinds of phones. 

\sys{}, on the other hand, addresses the scalability issues of outdoor environments through the virtual griding approach. It builds the fingerprint from the provider side. Hence, It needs only the phone-ID to localize users. It can provide accurate localization for different network types over all kinds of phones. In addition, it introduces spatial augmentation techniques to reduce the number of samples required to train a deep learning model. Moreover, its design includes a number of provisions to handle the noise in the input data as well as provide robust performance in different environments.

\section{Conclusion}
\label{conclusion}
In this paper, we proposed \sys{}: a provider-side fin-
gerprinting localization system that can provide high accuracy
localization for any cell phone. \sys{} leverages the provider cellular unlabeled data to build the fingerprint radio map. The radio map is then used to train a deep neural network model for low-end phones localization. Our system contains different modules to handle system challenges.   

We implemented and tested \sys{} in a typical realistic environment and compared its performance against existing systems. 
Results show a median accuracy of 29m, which outperforms the state-of-the-art cellular-based localization systems by more than 75.4\%.

\bibliographystyle{ACM-Reference-Format}
\bibliography{main}

\end{document}